\documentclass[aps,prl,twocolumn,nofootinbib,groupedaddress,superscriptaddress,secnumarabic,
,longbibliography]{revtex4-2}

\AtBeginDocument{
\heavyrulewidth=.08em
\lightrulewidth=.05em
\cmidrulewidth=.03em
\belowrulesep=.65ex
\belowbottomsep=0pt
\aboverulesep=.4ex
\abovetopsep=0pt
\cmidrulesep=\doublerulesep
\cmidrulekern=.5em
\defaultaddspace=.5em
}

\usepackage{float}
\usepackage{comment}
\usepackage{booktabs}
\usepackage[percent]{overpic}
\usepackage{microtype}
\usepackage{array}
\usepackage{physics}
\usepackage{graphicx}
\usepackage{bbold}
\usepackage[usenames,dvipsnames]{xcolor}
\usepackage{tikz}
\usepackage{epsfig}

\usepackage[export]{adjustbox}

\usepackage{amsmath,amssymb}
\usepackage{multirow}
\usepackage{bm}
\usepackage{mathtools}
\usepackage{dsfont}
\usepackage{amsfonts}
\usepackage[normalem]{ulem}
\usepackage{url}
\usepackage{wasysym}
\usepackage{multibib}
\newcites{SM}{Supplementary References} 

\usepackage[colorlinks=true,linkcolor=tum,citecolor=tum,urlcolor=tum, hypertexnames=false]{hyperref}
\usepackage{cleveref}

\def\ba#1\ea{\begin{align}#1\end{align}}
\def\bg#1\eg{\begin{gather}#1\end{gather}}
\def\bpm{\begin{pmatrix}}
\def\epm{\end{pmatrix}}

\definecolor{myBlue}{HTML}{0065BD}   
\newcommand{\tum}[1]{\textcolor{myBlue}{#1}}

\definecolor{ForestGreen}{HTML}{668000}
\definecolor{red1}{HTML}{FF4136}
\definecolor{green1}{HTML}{00802b}	
\definecolor{tum}{HTML}{0065bd}

\usetikzlibrary{angles,quotes}
\newcommand{\Ham}{\mathcal{H}}
\newcommand{\vk}{\mathbf{k}}
\newcommand{\Nuc}{{N_{\text{UC}}}}
\usepackage{tabulary}
\newcolumntype{K}[1]{>{\centering\arraybackslash}p{#1}}

\newcommand{\thomas}[1]{\textcolor{red}{\textbf{#1}}}

\newcommand{\eqnref}[1]{Eq.~(\ref{#1})}
\newcommand{\figref}[1]{Fig.~\ref{#1}}

\newcommand{\ourtitle}{Triangular $J_1$-$J_2$ Heisenberg Antiferromagnet in a Magnetic Field}

\allowdisplaybreaks

\begin{document}
\title{\textbf{\ourtitle}}

\author{T. Bader}
\affiliation{Technical University of Munich, TUM School of Natural Sciences, Physics Department, 85748 Garching, Germany}
\affiliation{Munich Center for Quantum Science and Technology (MCQST), Schellingstr. 4, 80799 M{\"u}nchen, Germany}
\author{S. Feng}
\affiliation{Technical University of Munich, TUM School of Natural Sciences, Physics Department, 85748 Garching, Germany}
\affiliation{Munich Center for Quantum Science and Technology (MCQST), Schellingstr. 4, 80799 M{\"u}nchen, Germany}
\author{S. Budaraju}
\affiliation{Technical University of Munich, TUM School of Natural Sciences, Physics Department, 85748 Garching, Germany}
\affiliation{Munich Center for Quantum Science and Technology (MCQST), Schellingstr. 4, 80799 M{\"u}nchen, Germany}
\author{F. Becca}
\affiliation{Dipartimento di Fisica, Universit\`a di Trieste, Strada Costiera 11, I-34151 Trieste, Italy}
\author{J. Knolle}
\affiliation{Technical University of Munich, TUM School of Natural Sciences, Physics Department, 85748 Garching, Germany}
\affiliation{Munich Center for Quantum Science and Technology (MCQST), Schellingstr. 4, 80799 M{\"u}nchen, Germany}
\affiliation{Blackett Laboratory, Imperial College London, London SW7 2AZ, United Kingdom}
\author{F. Pollmann}
\affiliation{Technical University of Munich, TUM School of Natural Sciences, Physics Department, 85748 Garching, Germany}
\affiliation{Munich Center for Quantum Science and Technology (MCQST), Schellingstr. 4, 80799 M{\"u}nchen, Germany}


\begin{abstract}
The behavior of the paradigmatic $J_1-J_2$ triangular lattice Heisenberg antiferromagnet in a magnetic field remains unsettled despite decades of study. We map out the phase diagram using three complementary approaches, including self-consistent nonlinear spin-wave theory, density-matrix renormalization group, and variational Monte Carlo. This combined analysis resolves the competition among different field-induced magnetic orders and magnetization plateaux across the classically frustrated parameter range. In particular, there is a finite range in the parameter regime around $J_2/J_1=\frac{1}{8}$ in which i) upon the application of the external field, the gapless quantum spin liquid acquires a finite density of monopoles, and ii) by further increasing the field, two plateaux are clearly obtained at $m=\frac{1}{3}$ and $m=\frac{1}{2}$.
We discuss the experimental importance of the consecutive magnetization plateaux transitions as a signature of an underlying quantum spin-liquid phase.  

\end{abstract}

\maketitle
\let\oldaddcontentsline\addcontentsline
\renewcommand{\addcontentsline}[3]{}


{\it \tum {Introduction--}} Triangular lattice antiferromagnets (TrAFM) are not only widely-used illustrations of geometrically frustrated magnets, but arguably the founding examples of an entire research field~\cite{lacroix2011introduction}. Famously, Wannier showed that the triangular Ising model realizes a critical liquid with an extensive classical ground state degeneracy~\cite{wannier1950antiferromagnetism} and the frustration is also manifest as an unusual step-like magnetization process of the model as a function of an applied magnetic field. Early on, it was shown that the magnetization jumps from zero to a plateau value $m=\frac{1}{3}$ and then to full saturation~\cite{metcalf1973phase}. Adding further next-nearest-neighbor (NNN) interactions $J_2$ in addition to the standard nearest-neighbor (NN) $J_1$ one enables a new magnetization plateau at $m=\frac{1}{2}$~\cite{metcalf1974ground,bradley2019robustness} from the interplay between geometric and exchange frustration. Beyond Ising models, the phase diagram is even richer for triangular Heisenberg-like systems with quantum fluctuations~\cite{nishimori1986magnetization,chubukov1991quantum} and thermal fluctuations stabilizing collinear plateaux phases~\cite{kawamura1985phase,gvozdikova2011magnetic,seabra2011phase}. In general, the study of magnetization plateaus has a long history~\cite{takigawa2010magnetization,starykh2015unusual,honecker2004magnetization} and is a telltale experimental signature of magnetic frustration~\cite{ono2003magnetization,shirata2012experimental}.  

The interest in the TrAFM has significantly increased in recent years because their competing interactions can stabilize sought-after quantum spin liquids (QSLs)~\cite{savary2016quantum,knolle2019field,broholm2020quantum} in a number of candidate materials \cite{kanoda2011mott,PhysRevB.92.041105,PhysRevB.100.241111,PhysRevLett.120.207203,PhysRevB.108.L220401, PhysRevB.109.014425,nature_SL1}. Since Anderson's original proposal~\cite{ANDERSON1973153}, the $S=\frac{1}{2}$ Heisenberg TrAFM has been a prominent theoretical candidate for realizing a QSL ground state. Indeed, upon the addition of NNN Heisenberg exchange of strength $0.07\lesssim J_2/J_1 \lesssim 0.16$ a QSL is by now well established numerically~\cite{PhysRevB.92.140403, PhysRevB.93.144411,PhysRevX.9.031026,PhysRevB.106.064428}. Recent progress on the experimental side has suggested that the Yb-based delafossites are expected to be accurately modeled by this type of Heisenberg model \cite{Scheie_2023,nature_CsYbSe2,PhysRevB.109.014425} but finding unambiguous signatures of QSL behavior remains a challenge~\cite{knolle2019field,scheie2024spectrum}. In this context, studying the phase diagram of TrAFM models beyond the QSL regime is important to gain theoretical insights into their stability and the degree to which basic models resemble physical materials.

In this work we study the phase diagram of the spin-$\frac{1}{2}$ Heisenberg TrAFM in a magnetic field described by the Hamiltonian
\begin{equation}
    \mathcal{H}=J_1\sum_{\langle i j \rangle} \vec{S}_i \cdot \vec{S}_j + J_2\sum_{\langle\langle i j \rangle\rangle} \vec{S}_i \cdot \vec{S}_j-H\sum_iS_i^z
\end{equation}
and concentrate on $0 \leq J_2/J_1 \leq 1$. The classical limit exhibits a three-sublattice phase for $0\leq J_2/J_1\leq \frac{1}{8}$, a four-sublattice phase for $\frac{1}{8}\leq J_2/J_1\leq 1$ and a fully polarized phase for $H\geq \max\{4.5J_1, 4(J_1+J_2)\}$, yet the ground state is infinitely degenerate in the former two (see supplemental material \cite{supp}).
Quantum fluctuations lift the degeneracy, which has been extensively studied and verified in the zero field limit \cite{PhysRevB.42.4800,PhysRevB.46.11137,PhysRevLett.60.2531}. Furthermore, the phase diagram for small and large $J_2$ including a magnetic field has been studied in linear spin wave theory (LSWT) with self-consistent non-linear spin wave theory (NLSWT) applied to the plateaux and Ginzburg-Landau theory on condensates \cite{PhysRevB.94.075136,PhysRevB.95.014425,PhysRevB.96.140406}. 

\begin{figure}[t]
\begin{center}
    \includegraphics[width=\linewidth]{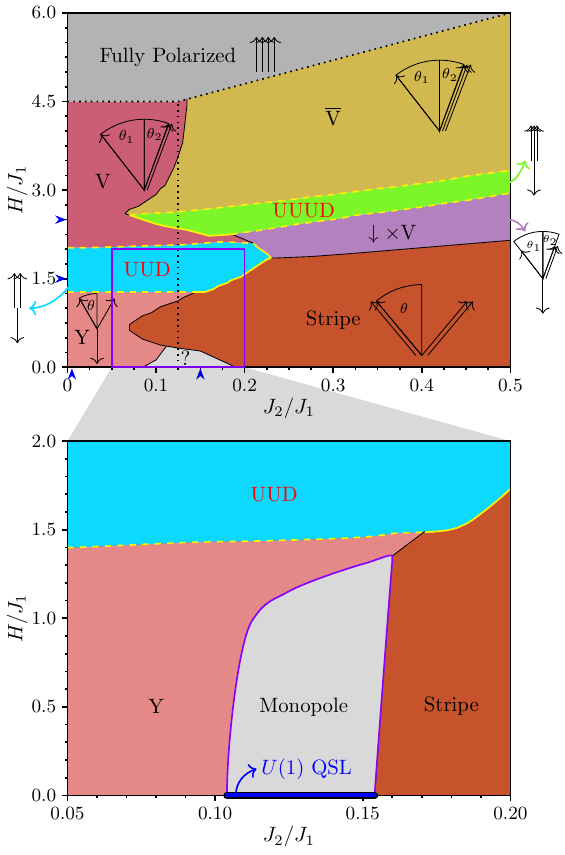}
    \caption{Top: Phase diagram of the $J_1-J_2-H$ model determined from self-consistent NLSWT on a $30\times 30$ triangular lattice with PBC. Dotted lines indicate classical phase boundaries between three- and four-sublattice and fully polarized phases. Solid (dashed) lines indicate first (second) order phase transitions. 
    The light gray area denotes the expected breakdown of semiclassical configurations as predicted from LSWT and mode condensation. The $m=\frac{1}{3}$ (UUD) and $m=\frac{1}{2}$ (UUUD) magnetization plateaux are highlighted in light blue and green, with their phase boundaries highlighted in yellow. The purple box contains a region beyond the reach of the semiclassical theory.
    Bottom: The low-field phase diagram of the $J_1-J_2-H$ model in the vicinity of QSL obtained from VMC on an $18 \times 18$ triangular lattice. 
    }
    \label{fig:phasediag}
\end{center}
\end{figure}

Despite the long research history, several open questions remain: i) the precise location of phase boundaries; ii) the extent of the magnetization plateaux around the interesting classical frustration point $J_2/J_1=\frac{1}{8}$; iii) the validity of semiclassical approaches; and, most importantly, iv) the fate of the critical QSL when switching on a magnetic field. Here, we determine the phase diagram by combining NLSWT, density-matrix renormalization group (DMRG) \cite{White1992,Hauschild2024}, and variational Monte Carlo (VMC) \cite{beccabook} analyses. Within NLSWT, we implement the Holstein–Primakoff (HP) transformation and extract self-consistent solutions including quantum corrections up to second order ($S^{-2}$) across the full range of $J_2$ and $H$ under study. We place these results alongside independent DMRG and VMC calculations on representative cuts through the phase space, using these complementary approaches to cross-validate one another and thereby establish the phase diagram shown in \figref{fig:phasediag}.
Remarkably, we find overlapping $m=\frac{1}{3}$ and $m=\frac{1}{2}$ magnetization plateaux above the zero-field QSL regime. Thus, the experimental observation of such consecutive plateaux is a strong indication that the Hamiltonian is in a parameter regime allowing for a zero field QSL. 
In addition, we show that applying a field to the critical QSL leads to a ground state with a finite density of monopoles; see \figref{fig:phasediag} for a quantitative estimate of the phase diagram in vicinity of the QSL regime.

\begin{figure*}[t]
    \includegraphics[width=\textwidth]{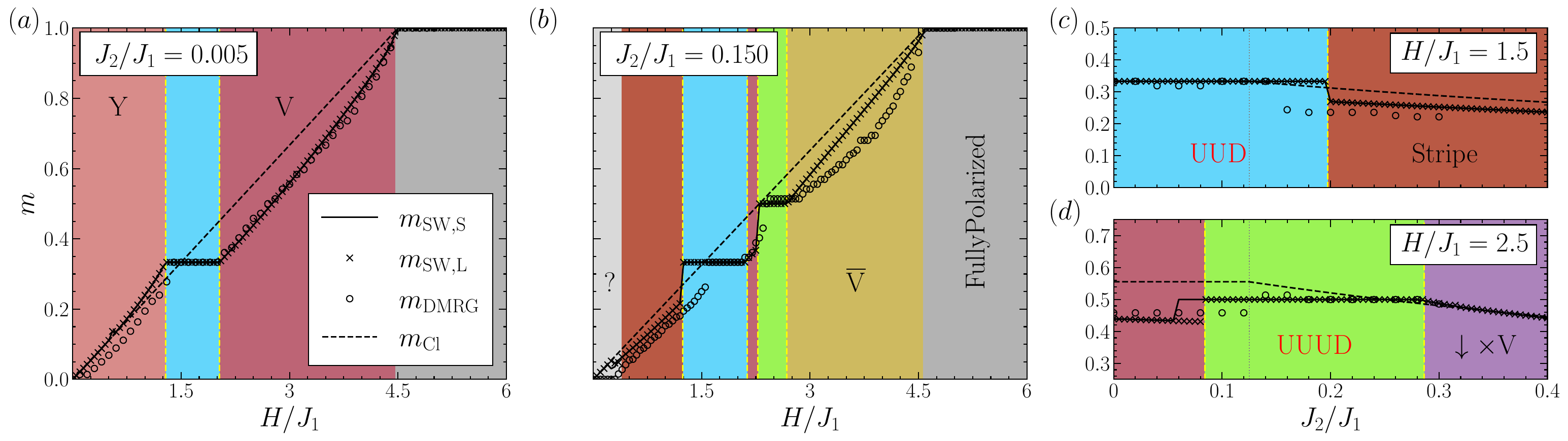}
    \caption{Magnetization cuts marked by the blue arrowheads on the axes of \figref{fig:phasediag}(top), obtained from different methods; phases are colored according to the results from self-consistent stable ($m_{SW,S}$) and local NLSWT ($m_{SW,S}$). The crosses are obtained from the local algorithm, the continuous line from the stable algorithm and the DMRG data obtained on a $6\times 24$ cylinder is shown using circles. Both NLSWT algorithms run on a $30\times 30$ lattice using PBC. The dashed line shows the classical magnetization. (a, b)  Plots as a function of $H/J_1$ at fixed $J_2/J_1$. (c, d) Plots of magnetization as a function of $J_2/J_1$ at different fixed magnetic fields $H=1.5$ and $H=2.5$. The horizontal dashed line indicates the classical phase transition. 
    }
    \label{fig:cuts}
\end{figure*}

{\it \tum {Semiclassical Theory.--} } We begin by outlining details of the well known HP transformation to obtain quantum corrections to classical ground states. The states considered are not collinear, hence the transformation is applied in a rotated frame $\vec{S}'$, chosen such that the $z$-axis is aligned with the respective classical spin direction. The spin inside the Hamiltonian is obtained using a standard rotation matrix $R$ such that $\vec{S}=R\vec{S}'$.
The transformation itself is given as
\begin{equation}
\label{eq:hp_gen}
    S'^+ = \sqrt{2S-a^\dagger a}a,~~
    S'^z = S-a^\dagger a,
\end{equation}
which we then expand for large $S$.

To compute the ground state energy, we include up to quartic terms in the expansion of the full Hamiltonian,  and treat quartic and cubic terms self-consistently. Since the ground state is (not always) collinear, we need to additionally consider terms linear in the operators. For classical ground states, the leading order vanishes; however, higher orders require a renormalization of angles such that any term linear in the bosonic creation and annihilation operators vanishes. For this work, we need the renormalized set of angles $\theta$ up to order $S^{-1}$, which is generally determined from the requirement
\begin{equation}
    \Ham^{(1)}(\theta)+\frac{1}{S}\overline{\Ham}^{(3)}(\theta)=0+\mathcal{O}(S^{-2}),
\end{equation}
combined with preserving the classical limit $\theta = \theta_{\text{cl}}+\mathcal{O}(S^{-1})$. The overline ($\overline{\Ham}$) indicates that terms have been contracted to scalar, linear and quadratic order in the boson operators. Solving this equation for $\theta$ yields the proper set of renormalized angles. We use the angle determined from these equations inside the Hamiltonian and perform the expansion thereof. In doing so, we do not expand the angle in $S^{-1}$ but rather treat it akin to a variational parameter
\begin{equation}
\label{eq:ham_exp}
\begin{split}
    \Ham (\theta)&= \Ham_{\text{cl}}(\theta)+\frac{1}{S}\Ham^{(2)}(\theta)+\frac{1}{S^2}\overline{\Ham}^{(4)}(\theta)\\
    &= NE_{\text{cl}}(\theta)+\frac{1}{2S}\sum_\vk \left(a_\vk^\dagger M_\vk(\theta) a_\vk-\Delta_\vk\right),
\end{split}
\end{equation}
where $N$ is the total number of lattice sites and the sum runs over all accessible points in the Brillouin zone. The ground state (GS) energy is then obtained by diagonalizing the Hamiltonian under periodic boundary conditions (PBC) using a standard Bogoliubov transformation.

To obtain a self-consistent solution, we make use of complementary two algorithms: The \textit{local algorithm} works at a single point in parameter space. At every step, it uses a linear combination of mean-field parameters from the previous two steps at some mixing angle and computes the ground state energy from \eqnref{eq:ham_exp} and new mean-field parameters from the eigenvalues thereof. This process is repeated until the energy has converged within bounds of $10^{-9}$ for ten consecutive steps.
Since this algorithm is prone to runaway effects and lattice-size-dependent instabilities, we use the \textit{stable algorithm} as a cross check. In this case, the local algorithm is performed at a point $J_2^{(0)}$, $H^{(0)}$ in parameter space, for which the ground state is well defined. The mean-field parameters of this solution are then used analogous to the local algorithm, however additionally modifying the parameters $J_2$ and $H$ in small increments at every step.
Clearly, the stable algorithm is preferable for computing the phase diagram (in the sense of computational cost) if the state is stable and adiabatically connected to the previous state. In contrast, the local algorithm is able to capture self-consistent ground states that may not be adiabatically connected to neighboring states in the sense of the MF parameters.

{\it \tum {Phase Diagram and Magnetization Plateaux.--}} 
The semiclassical analysis outlined above predicts the rich phase structure shown in the top panel of \figref{fig:phasediag}. 
At $H=0$, LSWT predicts a breakdown of the Y state, and hence any 120$^\circ$ ordered state, at $J_2/J_1 \gtrsim 0.085$ while the Stripe phase has no self-consistent solution without condensates below $J_2/J_1 \lesssim 0.16$, in agreement with the current literature. We thus find a finite region in parameter space connected to the QSL regime in which none of the classical states investigated are stable, as denoted by the purple box in the top panel of \figref{fig:phasediag}. The low-field phase diagram near the QMC regime is instead given by the VMC approach shown in the bottom panel of \figref{fig:phasediag}, whose details are to be discussed in the next section. 
Note, contrary to previous works~\cite{PhysRevB.95.014425,PhysRevB.96.140406}, we find that the stripe state is not the ground state close to the fully polarized phase for $J_2/J_1\geq \frac{1}{8}$ in self-consistent NLSWT. We also verified that the umbrella state, defined as an in-plane $120^\circ$ order with all spins canted along the direction of the magnetic field, is never the self-consistent solution in the phase diagram with lowest energy within our NLSWT. 

We now obtain a complete semiclassical characterization of the magnetic phases for $H \neq 0$ and variable next-nearest-neighbor exchange. In addition to the well-known field-induced UUD phase at small $J_2$, we find two consecutive magnetization plateaux at distinct magnetic fields: the UUD plateau with $m = \frac{1}{3}$ for small-to-moderate NNN exchange $0 < J_2/J_1 \lesssim \frac{1}{4}$ at moderate fields (blue area in Fig.~\ref{fig:cuts}a), and the UUUD plateau with $m = \frac{1}{2}$ for $J_2/J_1 \gtrsim 0.08$ at slightly higher fields (green area in Fig.~\ref{fig:cuts}b). Both plateaux remain stable beyond the classical regime in $H$ and $J_2$ once non-linear corrections are included. Remarkably, as clearly shown in Fig.~\ref{fig:cuts}b, there is an overlapping window around $J_2/J_1 \sim \frac{1}{8}$ in which both UUD and UUUD phases appear successively upon increasing the field, while a QSL is expected in the zero-field limit. This two-plateau structure therefore provides a useful diagnostic of the range of NNN Heisenberg couplings where a Dirac QSL is likely to occur at low fields, albeit its robustness against anisotropies remains an open question for future work. 

The consecutive magnetization plateaux could be accessible experimentally not only through chemical tuning of the exchange ratio but also via application of pressure or strain.
Similar tuning has been achieved on Kitaev and Heisenberg interactions on the honeycomb lattice~\cite{PhysRevB.98.121107,Sakrikar_2025}. 
In recent experiments on the TrAFM~\cite{PhysRevB.109.014425,belbase2025finitespinondensityofstatestriangularlattice}, the $m=\frac{1}{3}$ plateau has been observed while the $m=\frac{1}{2}$ plateau was absent for candidate materials NaYbSe$_2$ and TlYbSe$_2$ which have $J_2/J_1$ ratios for which we predict the $m=\frac{1}{2}$ plateau to exist. It points to the relevance of other non-Heisenberg interactions which can destabilize the plateaux. For example, an easy-plane anisotropy of the NNN exchange interaction energetically favors the canted stripe state compared to the UUUD plateau. Such XXZ-type anisotropies have been shown to persist beyond classical approximations in the $J_2=0$ sector \cite{PhysRevB.91.081104}. Single-ion easy-axis anisotropies have also been shown to impact the extent of plateaux's~\cite{PhysRevB.83.134412}. Furthermore, Dzyaloshinskii-Moriya or spin-scalar-chirality type interactions may destabilize plateaus as well.

Although NLSWT offers a controlled semiclassical description of the competing ordered phases, we subject its predictions to a stricter test using unbiased DMRG. The DMRG calculations, while necessarily performed on finite cylinders and therefore subject to finite-size constraints, fully incorporate quantum fluctuations and produce magnetization curves that remain consistent with the semiclassical phase structure across all accessible system sizes.
Specifically, we perform grand-canonical DMRG (i.e. without fixing the total magnetization) on a $6 \times 24$ cylinder with fixed bond dimension $\chi = 1600$ and controlled truncation error $\sim 10^{-7}$, thus obtaining the optimal magnetization $m$ for each field $H$. 
Calculations along several cuts of the phase diagram are  shown in \figref{fig:cuts}. 
The resulting magnetization curve is then directly compared to the magnetization obtained in NLSWT:
\begin{equation}
    m = \frac{1}{S\Nuc }\sum_{i=1}^\Nuc \cos(\theta_i)(S-n_i)
\end{equation}
where $n_i=\expval*{a_i^\dagger a_i}$ and $\Nuc$ are respectively the boson occupation number and the number of sites per unit cell.

Figure~\ref{fig:cuts}a shows the cut in $H$ with a fixed small $J_2/J_1 = 0.005$, where we find quantitative agreement between both algorithms of NLSWT and DMRG simulations in all magnetic phases. 
For larger $J_2/J_1 = 0.15$, shown in \figref{fig:cuts}b, the same progression of phases is obtained in both methods; remarkably, both methods predict two successive magnetization plateaux (blue and green), albeit with minor deviations in the precise location of phase transition. 
It should be noted that these minor deviations lie close to the classical phase transition at $J_2/J_1=\frac{1}{8}$, where we observe mean-field parameters like $n_i$ becoming larger, thus worsening the predictive power of NLSWT. This is visible in Figs.~\ref{fig:cuts}c and d, cuts along $J_2/J_1$ at fixed magnetic fields $H = 1.5$ and $H = 2.5$ respectively, where the quantitative agreement between NLSWT and DMRG breaks down close to $J_2/J_1=\frac{1}{8}$ [dotted lines in Figs.~\ref{fig:cuts}c, d]. 
Despite the perturbative nature of NLSWT and the finite-size effects of DMRG, the close agreement between these complementary methods lends strong support to the magnetic phases we identify. By contrast, in the non-magnetic QSL regime in the zero-field limit (purple box in Fig.~\ref{fig:phasediag}), NLSWT breaks down completely.

\begin{figure}
    \includegraphics[width=\linewidth]{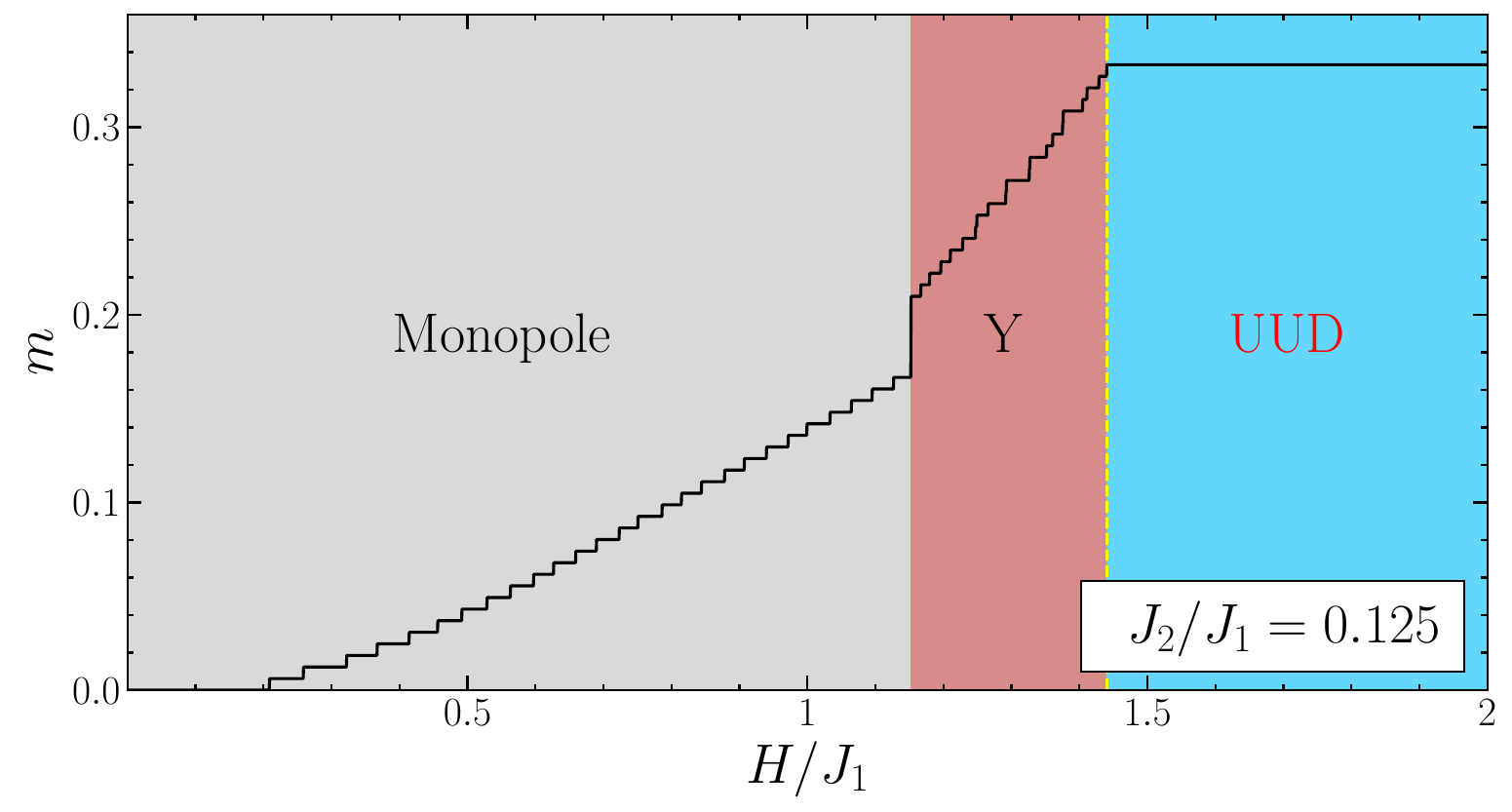}
    \caption{ Low-field magnetization curve at $J_2/J_1=\frac{1}{8}$ obtained via VMC calculations performed on the $18\times18$ triangular lattice with PBC. The plateau at $m=0$ is due to the finite-size gap of the $Q=1$ monopole excitation. A clear first order transition from the monopole to the Y phase is observed, and the latter continuously evolves into the UUD. } 
    \label{fig:vmc_mag_curve}
\end{figure}

{\it \tum {VMC analysis in the vicinity of QSL phase--}}
To probe the frustrated regime near $J_{2}/J_{1} = \frac{1}{8}$, we employ a VMC approach based on Gutzwiller-projected fermionic wavefunctions representing the relevant competing orders. For each fixed $S^{z}$ sector, from $0$ up to $m = \frac{1}{3}$, we construct and optimize the corresponding ansätze (e.g., canted-stripe, umbrella, Y, and monopole states) and compare their variational energies to assess the dominant correlations. The unprojected fermionic hamiltonian has the general form $\mathcal{H}_{\rm aux} = \sum_{\langle ij\rangle,\sigma} \chi_{ij} c_{i\sigma}^\dagger c_{j\sigma} + {\rm H.c.} + \sum_i \vec{M}_i \cdot \vec{S}_i $, where the parameters $\chi_{ij}$ and $\vec{M}_i$ differ for each phase
(details to be provided in \cite{unpub}). In particular, the monopole states (corresponding to finite densities of the gapless monopole excitations of the Dirac spin liquid~\cite{song2019unifying}) have $\vec{M}_i = 0$ and are generated by threading an internal gauge flux $2\pi Q$ uniformly through all triangles \cite{sasank25mono} and occupying the resulting $2Q$ fermionic zero modes with up spins, yielding a unique state with $S^{z}=Q$. The corresponding magnetization curve at $J_{2}/J_1=\frac{1}{8}$ is shown in \figref{fig:vmc_mag_curve}. The projected monopole states, despite having no tunable parameters, maintain the lowest variational energy among all competing ansätze up to $H/J_1 \approx 1.15$ ($m \approx \frac{1}{6}$), beyond which a sharp first-order transition occurs into the Y phase. The VMC-based estimate of the phase boundaries is provided in the bottom panel of \figref{fig:phasediag}.
These results demonstrate that in the vicinity of the QSL, the ground state in the presence of a field is a state with a finite density of monopoles. A detailed characterization of this phase is given in a companion work~\cite{unpub}. 

{\it \tum {Discussion and Conclusion.--} } 
We obtained the phase diagram of the $J_{1}-J_{2}-H$ triangular lattice Heisenberg antiferromagnet by integrating three complementary methods: self-consistent NLSWT (including all $S^{-2}$ corrections), large-scale DMRG, and variational Monte Carlo. NLSWT provides global structure and identifies candidate phases; DMRG supplies an unbiased quantum benchmark along representative cuts; VMC resolves the regime where semiclassical methods fail—namely the vicinity of the classical frustration point $J_{2}/J_{1}=\frac{1}{8}$, where the zero-field Dirac QSL emerges. Taken together, the methods yield a coherent and internally consistent phase diagram. First, non-linear quantum fluctuations substantially enlarge both the UUD ($m=\frac{1}{3}$) and UUUD ($m=\frac{1}{2}$) plateaux in both $H$ and $J_2/J_1$. Second, these plateaux overlap in the parameter window connected to the zero-field QSL: increasing the field produces two successive magnetization plateaux at $m=\frac{1}{3}$ and $m=\frac{1}{2}$, a distinctive experimental fingerprint of a system tuned close to the QSL regime. Third, away from phase boundaries and the QSL region, NLSWT with fully renormalized angles quantitatively reproduces the DMRG results, confirming its reliability over most of the phase diagram. 
Finally, VMC shows that applying a field onto a Dirac QSL leads to a finite-density monopole phase that persists up to a first-order transition into the Y state. Its detailed characterization is addressed in a companion work~\cite{unpub}.

\acknowledgments{
{\it Acknowledgments.-- }
We acknowledge helpful discussions and related collaborations with Markus Drescher, Francesco Ferrari, Roderich Moessner, Urban Seifert, Roser Valenti, and Josef Willsher, as well as Arnab Banerjee, Allen Scheie, Alan Tennant, and Alexander Tsirlin. We thank Anna Keselmann and Oleg Starykh for helpful discussions. 
We acknowledge support from the Deutsche Forschungsgemeinschaft (DFG, German Research Foundation) under Germany’s Excellence Strategy (EXC–2111–390814868 and ct.qmat EXC-2147-390858490), and DFG Grants
No. KN1254/1-2, KN1254/2-1 TRR 360 – 492547816 [14] and SFB 1143 (project-id 247310070), as well as the Munich Quantum Valley, which is supported by the Bavarian state government with funds from the Hightech Agenda Bayern Plus. J.K. further acknowledges support from the Imperial-TUM flagship partnership.  
}

{\it Note added.--} 
We learned of an independent complementary study \cite{keselman2025} upon completing this work, which reports a similar phase diagram.

\bibliography{biblio}

\clearpage
\appendix
\begin{widetext}
\begin{center}
\textbf{\large Supplementary Information for \\ ``\ourtitle"}

\medskip
T. Bader$^{1,2}$, S. Feng$^{1,2}$, S. Budaraju$^{1,2}$, F. Becca$^{3}$, J. Knolle$^{1,2,4}$, F. Pollmann$^{1,2}$

\medskip
{\it ${}^1$Technical University of Munich, TUM School of Natural Sciences, Physics Department, 85748 Garching, Germany}

{\it ${}^2$Munich Center for Quantum Science and Technology (MCQST), Schellingstr. 4, 80799 M{\"u}nchen, Germany}

{\it ${}^3$Dipartimento di Fisica, Universit`a di Trieste, Strada Costiera 11, I-34151 Trieste, Italy}

{\it ${}^4$Blackett Laboratory, Imperial College London, London SW7 2AZ, United Kingdom}

\end{center}
\end{widetext}
\makeatletter
\@removefromreset{equation}{section}
\makeatother

\setcounter{equation}{0}
\setcounter{figure}{0}
\setcounter{table}{0}
\setcounter{page}{1}
\setcounter{section}{0}
\makeatletter
\renewcommand{\theequation}{S\arabic{equation}}
\renewcommand{\thefigure}{S\arabic{figure}}
\renewcommand{\bibnumfmt}[1]{[S#1]}

\setcounter{secnumdepth}{3}  

\makeatletter
\def\thesection{S\arabic{section}}
\def\thesubsection{S\arabic{section}.\arabic{subsection}}
\def\thesubsubsection{S\arabic{section}.\arabic{subsection}.\arabic{subsubsection}}
\makeatother
\let\addcontentsline\oldaddcontentsline
\tableofcontents

\section{General Framework}

Our goal is to compute quantum corrections and magnon dispersion around classical ground states of the $J_1-J_2-H$ model on the triangular lattice, defined by the Hamiltonian
\begin{equation}
	\Ham = \frac{J_1}{S^2} \sum_{\langle i j \rangle} \vec{S}_i \cdot \vec{S}_j+ \frac{J_2}{S^2} \sum_{\langle\langle i j \rangle\rangle} \vec{S}_i \cdot \vec{S}_j-\frac{H}{S} \sum_i S_i^z.
\end{equation}
We have normalized all interactions by the spin quantum number, such that the classical phase transitions are independent of $S=\abs*{\vec{S}_i}$.\par
Expanding around a classical state amounts to performing an expansion around large $S$, i.e. an expansion in $S^{-1}$. For this, we shall use the Holstein Primakoff method, where a spin, classically aligned in the positive $z$-direction is expressed in terms of bosonic operators $a, a^\dagger$ as
\begin{equation}
\label{eq:hp_gen2}
	\begin{split}
		S^+ =& \sqrt{2S-a^\dagger a}a\\
		S^z =& S-a^\dagger a,
	\end{split}
\end{equation}
such that $\expval{\vec{S_i}} = S \vec{e}_z + \mathcal{O}(S^{-\frac{1}{2}})$. For an arbitrarily rotated classical spin, determined by the polar and azimuthal angles $\theta, \phi$, the expansion is performed in a rotated, local frame, such that the spin in the laboratory frame is given by
\begin{equation}
	\vec{S}=R_\phi R_\theta \vec{S}', 
\end{equation}
leading to the very general expression
\begin{equation}
\label{eq:spin_def}
	\left(\begin{array}{c}
		S^x \\ S^y \\ S^z 
	\end{array}\right)= \left(\begin{array}{c}
	c_\theta c_\phi{S'}^x - s_\phi{S'}^y+ s_\theta c_\phi{S'}^z\\ s_\phi ({S'}^x c_\theta+{S'}^z s_\theta)+{S'}^y c_\phi\\{S'}^z c_\theta-{S'}^x s_\theta
	\end{array}\right),
\end{equation}
where we plug in the expansion from \eqnref{eq:hp_gen2} for the components $S'$. Using an ordinary Taylor expansion, we thus obtain an expansion in $S^{-1}$ (or, more precisely, $S^{-\frac{1}{2}}$).

\section{Classical Ground States}

The classical ground states of the system are given in terms of a three- or four-site unit cell for $0\leq J_2 \leq \frac{1}{8}$ and $\frac{1}{8} \leq J_2 \leq 1$, respectively.

\subsection{Three-Site Unit Cell}

For $\Nuc=3$, the classical Hamiltonian may be rewritten as
\begin{equation}
\begin{split}
	\Ham_{\text{cl}} = \frac{J_1N_p}{4S^2}&\left[\left(\vec{S}_{1}+\vec{S}_{2}+\vec{S}_{3}-\frac{SH\vec{e}_z}{3J_1}\right)^2\right.\\&\left.+\frac{6S^2J_2}{J_1}-3S^2-\frac{S^2H^2}{9J_1^2}\right],
\end{split}
\end{equation}
$N_p$ is the number of triangular plaquettes. We can immediately see that, under the assumption $\Nuc=3$, any state with
\begin{equation}
	\vec{S}_{1}+\vec{S}_{2}+\vec{S}_{3}=\frac{SH\vec{e}_z}{3J_1}
	\label{eq:gs3_def}
\end{equation}
is a ground state, as long as this condition can be met, i.e., $H\leq 9J_1$. For $H>9J_1$, the ground state is given by the fully polarized state with $\vec{S}_i = S\vec{e}_z$. The ground state energy per site is given by
\begin{equation}
	\frac{E_{\text{gs,cl}}}{N}=\left\{\begin{array}{ll}3J_2-\frac{3J_1}{2}-\frac{H^2}{18J_1}&H\leq 9J_1,\\ 3J_1+3J_2-H & H>9J_1.\end{array}\right.
\end{equation}
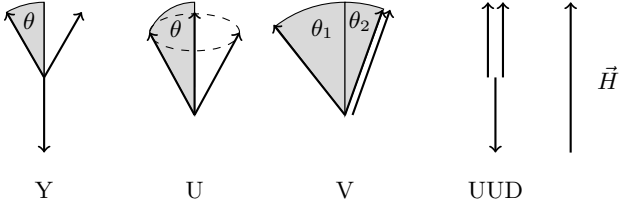
\begin{figure}[H]
	\begin{center}
\begin{tikzpicture}
	\draw [fill=gray!30] (1,0) -- (1,1) arc(90:120:1) -- cycle;
	\draw [fill=gray!30] (3,-.5) -- (3,1) arc(90:157:.65) -- cycle;
	\draw [fill=gray!30] (5,-.5) -- (5,1) arc(90:128:1.5) -- cycle;
	\draw [fill=gray!30] (5,-.5) -- (5,1) arc(90:70:1.5) -- cycle;
	\draw[dashed] (3,.6) ellipse (17pt and 7pt);
	\node (test) at (.8,.75) {$\theta$};
	\node (test) at (2.75, .65) {$\theta$};
	\node (test) at (4.7,.65) {$\theta_1$};2
	\node (test) at (5.2,.75) {$\theta_2$};
	\node (test) at (8.5, 0) {$\vec{H}$};
	\node (test) at (1, -1.5) {Y};
	\node (test) at (3, -1.5) {U};
	\node (test) at (5, -1.5) {V};
	\node (test) at (7, -1.5) {UUD};
	\draw[black, thick, ->] (1,0) -- (1,-1);
	\draw[black, thick, ->] (1,0) -- (0.5,0.87);
	\draw[black, thick, ->] (1,0) -- (1.5,0.87);
	\draw[black] (1,0)-- (1,1);
	\draw[black, thick, ->] (3,-.5) -- (3,.87);
	\draw[black, thick, ->] (3,-.5) -- (2.4,.6);
	\draw[black, thick, ->] (3,-.5) -- (3.6,.6);
	\draw[black] (3,-.13) -- (3,1);
	\draw[black, thick, ->] (5,-.5) -- (4.05,.7);
	\draw[black, thick, ->] (5,-.5) -- (5.5,.9);
	\draw[black, thick, ->] (5.1,-.5) -- (5.6,.9);
	\draw[black, thick, ->] (6.9,0) -- (6.9,1);
	\draw[black, thick, ->] (7.1,0) -- (7.1,1);
	\draw[black, thick, ->] (7,0) -- (7,-1);
	\draw[black, thick, ->] (8,-1) -- (8,1);
\end{tikzpicture}
	\end{center}
	\caption{The four three-sublattice states investigated. Note that the umbrella (U) state is non-coplanar compared to Y, V and UUD state.}
	\label{fig:three_sublat_states}
\end{figure}
The full ground state manifold is too large to be entirely investigated in this work, hence we will focus on four classical configurations shown in \cref{fig:three_sublat_states}. The angles are chosen such that the defining property \eqnref{eq:gs3_def} is fulfilled.

\subsection{Four-Site Unit Cell}

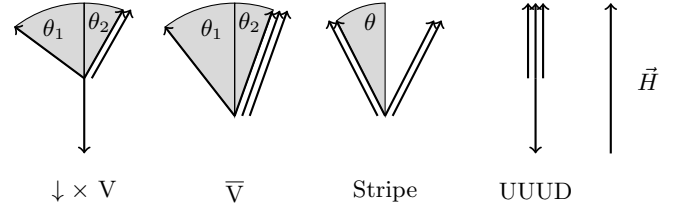
\begin{figure}[H]
	\begin{center}
		\begin{tikzpicture}
			\draw [fill=gray!30] (5,-.5) -- (5,1) arc(90:130:1) -- cycle;
			\draw [fill=gray!30] (1,0) -- (1,1) arc(90:128:1.5) -- cycle;
			\draw [fill=gray!30] (1,0) -- (1,1) arc(90:70:1.5) -- cycle;
			\draw [fill=gray!30] (3,-.5) -- (3,1) arc(90:128:1.5) -- cycle;
			\draw [fill=gray!30] (3,-.5) -- (3,1) arc(90:70:1.5) -- cycle;
			\node (test) at (.6,.65) {$\theta_1$};
			\node (test) at (1.2,.75) {$\theta_2$};
			\node (test) at (2.7,.65) {$\theta_1$};
			\node (test) at (3.2,.75) {$\theta_2$};
			\node (test) at (4.8,.75) {$\theta$};
			\node (test) at (8.5, 0) {$\vec{H}$};
			\node (test) at (1, -1.5) {$\downarrow \times$ V};
			\node (test) at (3, -1.5) {$\overline{\text{V}}$};
			\node (test) at (5, -1.5) {Stripe};
			\node (test) at (7, -1.5) {UUUD};
			\draw[black, thick, ->] (1,0) -- (1,-1);
			\draw[black, thick, ->] (1,0) -- (0.07,0.69);
			\draw[black, thick, ->] (1,0) -- (1.52,0.915);
			\draw[black, thick, ->] (1.1,0) -- (1.62,0.915);
			\draw[black] (1,0)-- (1,1);
			\draw[black, thick, ->] (3,-.5) -- (2.07,.69);
			\draw[black, thick, ->] (3,-.5) -- (3.5,.9);
			\draw[black, thick, ->] (3.1,-.5) -- (3.6,.9);
			\draw[black, thick, ->] (3.2,-.5) -- (3.7,.9);
			\draw[black] (3,-.5) -- (3,1);
			\draw[black, thick, ->] (4.9,-.5) -- (4.25,.77);
			\draw[black, thick, ->] (5,-.5) -- (4.35,.77);
			\draw[black, thick, ->] (5,-.5) -- (5.65,.77);
			\draw[black, thick, ->] (5.1,-.5) -- (5.75,.77);
			\draw[black, thick, ->] (6.9,0) -- (6.9,1);
			\draw[black, thick, ->] (7.1,0) -- (7.1,1);
			\draw[black, thick, ->] (7,0) -- (7,1);
			\draw[black, thick, ->] (7,0) -- (7,-1);
			\draw[black, thick, ->] (8,-1) -- (8,1);
		\end{tikzpicture}
	\end{center}
	\caption{The four four-sublattice states investigated.}
	\label{fig:four_sublat_states}
\end{figure}

In the four-sublattice phase, the classical Hamiltonian may be rewritten as 
\begin{equation}
	\Ham =  \frac{NJ}{4S^2}\left[\left(\sum_{i=1}^4\vec{S}_i-\frac{SH\vec{e}_z}{2J}\right)^2-\frac{S^2(H^2+16J^2)}{4J^2}\right],
\end{equation}
where $J=J_1+J_2$. Similar to the tree-site unit cell, any configuration with $\sum_{i=1}^4 \vec{S}_i=\frac{SH\vec{e}_z}{2J}$ is a ground state, as long as $H\leq 8J$ and the fully polarized state for $H>8J$. Hence the ground state energy per site is given by
\begin{equation}
	\frac{E_{\text{gs,cl}}}{N}=\left\{\begin{array}{ll}-\frac{H^2+16J^2}{16J}&H\leq 8J,\\ 3J-H & H>8J.\end{array}\right..
\end{equation}
Again, the classical ground state manifold is massively degenerate. We here consider the four states shown in \cref{fig:four_sublat_states}.

\section{Expansion in $S^{-1}$}

\subsection{Outline}

The main goal of our computation is to obtain the proper ground state within the framework of Holstein Primakoff, i.e., the ground state up to some order in $S^{-1}$. There is one caviat: We cannot perform an exact computation at an arbitrary order in the expansion. However, we can treat all terms beyond $S^{-1}$ in a mean-field manner such that at most quadratic terms remain.\par
To obtain the ground state energy, we first need to plug in the Holstein Primakoff representation of the spins into the Hamiltonian and expand in $S^{-1}$
\begin{equation}
	\Ham(\Theta) = \sum_{n=0}^2S^{-n/2}\Ham^{(n)}(\Theta)+\sum_{n=3}^\infty S^{-n/2}\overline{\Ham}^{(n)}(\Theta),
	\label{eq:Ham_exp_1}
\end{equation}
where an overline indicates treatment in mean-field theory until the Hamiltonian contribution is at most quadratic. The Hamiltonian in this case is a function of the set of classical angles $\Theta$ that we can, in principle, choose arbitrarily.\par
Now, two things are needed for the full self-consistent ground state: First, we need to know the \textit{renormalization} of the angles such that all terms linear in a bosonic operator in \eqnref{eq:Ham_exp_1} vanish. Second, we need to evaluate the ground state energy given a certain set of angles. 

\subsection{Angle Renormalization}

The angle renormalization is straightforward. We choose an initial configuration $\Theta^{(0)}$ defined by
\begin{equation}
	\Ham^{(1)}(\Theta^{(0)})=0,
\end{equation}
which is the same as requiring $\Theta^{(0)}$ is a classical ground state configuration. The renormalized angles
\begin{equation}
	\Theta^{[2n]}= \sum_{m=0}^n \frac{\Theta^{(2m)}}{S^m}
\end{equation}
are determined by requiring
\begin{equation}
	\Ham^{(1)}(\Theta^{[2n]})+\sum_{m=1}^{n}\frac{\overline{\Ham}^{(2m+1)}(\Theta^{[2n]})}{S^m}=0+\mathcal{O}(S^{-n-1}).
\end{equation}
Our goal is to obtain the ground state energy up do $\mathcal{O}(S^{-2})$, hence $\Theta^{(4)}$ can only enter through the classical energy contribution. However since $\Theta^{(0)}$ is a classical ground state configuration, $\Theta^{(4)}$ is irrelevant for our purposes as linear deformations in the angles lead to at least quadratic deformations in the classical energy. Hence for our purposes, the equation
\begin{equation}
	\Ham^{(1)}(\Theta^{[2]})+\frac{\overline{\Ham}^{(3)}(\Theta^{[2]})}{S}=0+\mathcal{O}(S^{-2}).
\end{equation}
is sufficient to obtain the renormalized angles.

\subsection{Quantum Ground State Energy}

\subsubsection{Matrix Representation of the Hamiltonian}

With the angles being properly renormalized, we are left only with quadratic bosonic terms from \eqnref{eq:Ham_exp_1}, i.e. we can readily write down the Hamiltonian up to order $\mathcal{O}(S^{-3})$
\begin{equation}
\begin{split}
	\Ham(\Theta)=&\Ham^{(0)}(\Theta^{[2]})+\frac{1}{S}\Ham^{(2)}(\Theta^{[2]})\\&+\frac{1}{S^2}\overline{\Ham}^{(4)}(\Theta^{[0]})+\mathcal{O}(S^{-3}).
	\label{eq:Ham_exp_2}
\end{split}
\end{equation}
Note that the above expansion is not the only possible expansion up to this order. In fact, it might seem unintuitive to not expand every term explicitly, e.g. 
\begin{equation}
	\Ham^{(0)}(\Theta^{[2]}) = \Ham^{(0)}(\Theta^{(0)})+\frac{(\Theta^{(2)}\cdot \nabla)^2\Ham^{(0)}(\Theta^{(0)})}{2S^2}.
	\label{eq:Ham_exp_3}
\end{equation}
The reason for keeping the expansion the way of \eqnref{eq:Ham_exp_2} is that it is numerically much more stable. The key lies in the fact that an expansion like
\begin{equation}
	\cos(\phi+\delta \phi) = \cos(\phi)-\delta \phi \sin(\phi)+\mathcal{O}(\delta \phi^2)
\end{equation}
disrupts the boundedness of the cosine, $\cos(\phi+\delta \phi)\in [-1,1]$ while $\cos(\phi)-\delta \phi \sin(\phi)\in (-\infty, \infty)$. Physically, this is reasonable, as an expansion like \eqnref{eq:Ham_exp_3} could lead to Heisenberg interactions to be renormalized beyond the spectrum of the operator.\par
Let us now turn in more detail to the treatment of the different terms in \eqnref{eq:Ham_exp_2}. $\Ham^{(0)}(\Theta^{[2]})=NE_{\text{cl}}(\Theta^{[2]})$ is simply the classical energy evaluated at the renormalized angles, while $\Ham^{(2)}$ and $\overline{\Ham}^{(4)}$ are the first and second order quantum corrections. The former contains only terms quadratic in $a_i, a_i^\dagger$ while the latter includes quadratic contributions as well as constants from the mean-field decoupling.\par
We may rewrite $\Ham^{(2)}$ and $\overline{\Ham}^{(4)}$ by transforming to momentum space and using 
\begin{equation}
	a_\vk = (a_{1,\vk},\dots, a_{\Nuc, \vk}, a_{1,-\vk}^\dagger, \dots, a_{\Nuc, -\vk}^\dagger)^\text{T}.
\end{equation}
This leads to an expression of the form
\begin{equation}
	\Ham^{(2)}+\frac{1}{S}\overline{\Ham}^{(4)}= \frac{1}{2}\sum_\vk a_\vk^\dagger M_\vk a_\vk - \Delta_\vk
	\label{eq:Ham_M_1}
\end{equation}
where we drop the arguments from now on and assume the renormalized angle is used to the appropriate order. In \eqnref{eq:Ham_M_1}, $M_\vk$ is a matrix of the form
\begin{equation}
	M_\vk = \left(\begin{array}{cc}
A_\vk & B_\vk \\ B_{-\vk}^\ast & A_{-\vk}^\ast	
\end{array}
\right) = \left(\begin{array}{cc}
A_\vk & B_\vk \\ B_\vk & A_{\vk}
\end{array}
\right),
\label{eq:Ham_M_2}
\end{equation}
with $A_\vk$ determined form $a_{i,\vk}^\dagger a_{j,\vk}$ terms and $B_\vk$ from $a_{i,\vk}^\dagger a_{j,\vk}^\dagger$ terms. $\Delta_\vk$ is the vacuum energy. At order $S^{-1}$, $\Delta_\vk=\tr{A_\vk}$, stemming from the fact that the bottom right $A_\vk$ matrix in \eqnref{eq:Ham_M_2} creates non-normal ordered terms $a_{i,-\vk} a_{j,-\vk}^\dagger$. At order $S^{-2}$, $\Delta_{\vk}$ also contains all terms from the mean-field contractions. Note furthermore that we assumed 
\begin{equation}
	A_{-\vk}^\ast = A_\vk, \quad B_{-\vk}^\ast = B_{\vk}.
	\label{eq:Ham_M_3}
\end{equation}
For coplanar configurations, this is true at order $\mathcal{O}(S^{-1})$ as there is a residual $SO(2)$ symmetry in the definition of \eqnref{eq:spin_def} related to the choice of plane. We can use this freedom to keep $S^{y}=S'^{y}$, such that there are only real coefficients in front of any operator in the expansion in position space. This implies that any imaginary contribution in the matrix \eqnref{eq:Ham_M_2} originates from the connections $e^{i\vk \cdot \mathbf{d}}$ from transformation to momentum space. We will later show that this leads to the statement being true at order $\mathcal{O}(S^{-2})$. 

\subsubsection{Bogoliubov Transform}

To diagonalize the Hamiltonian, we will perform a Bogoliubov transform. That is, we introduce matrices $T_\vk$ and 
\begin{equation}
	Y = \left(\begin{array}{cc}
\mathbf{1}_\Nuc & 0 \\ 0 & -\mathbf{1}_\Nuc	
\end{array}
\right),
\end{equation}
with
\begin{equation}
	T_\vk^\dagger Y T_\vk = Y\quad \text{and}\quad T_\vk^\dagger M_\vk T_\vk=\Omega_\vk,
\end{equation}
where
\begin{equation}
	\Omega_\vk = \left(\begin{array}{cc}
\omega_\vk & 0 \\ 0 & \omega_\vk
\end{array}
\right)
\end{equation}
contains the diagonal matrices $\omega_\vk$ with the magnon energies on it's diagonal, $\omega_\vk = \text{diag}(\omega_{1,\vk},\dots, \omega_{\Nuc, \vk})$. Furthermore, we define new operators
\begin{equation}
	b_\vk = T_\vk^{-1}a_\vk = YT_\vk^\dagger Y a_\vk,
\end{equation}
where the symplecticity of $T_\vk$ preserves the bosonic commutation relations.\par
In practice, we cannot obtain a numerically feasable closed form expression for the matrix $T_\vk$, hence we have to take a different route: Note that
\begin{equation}
	Y \Omega_\vk = Y T_\vk^\dagger Y^2 M_\vk T_\vk = T_\vk^{-1} YM_\vk T_\vk,
\end{equation}
we see that $T_\vk$ is a diagonalizing matrix of $YM_\vk$. Note additionally that
\begin{equation}
	YM_\vk T_{\vk}\vec{e}_i =T_\vk T_\vk^{-1} YM_\vk T_{\vk}\vec{e}_i = \pm \omega_{i,\vk} T_\vk \vec{e}_i,
\end{equation}
i.e., $T_\vk \vec{e}_i$ is an eigenvector of $Y M_\vk$. If the eigenvalues are non-degenerate (that is $\omega_{i,\vk} \neq \omega_{j,\vk} \neq 0\,\forall i\neq j$), the set of eigenvectors is unique up to normalization. The condition of symplecticity translates to the normalization
\begin{equation}
	v_{i,\vk}^\dagger Y v_{i,\vk}=\pm 1.
\end{equation}

\subsubsection{Ground State Energy}

As we now know $T_\vk$ and $\omega_{i,\vk}$ we can evaluate the ground state energy. It originates from bringing the Hamiltonian \eqnref{eq:Ham_M_1} into normal order, generating the ground state energy as
\begin{equation}
	\expval{\Ham^{(2)}+\frac{1}{S}\overline{\Ham}^{(4)}}=\frac{1}{2}\sum_\vk \sum_{i}\omega_{i,\vk}-\Delta_\vk
\end{equation}
Furthermore, we can determine mean-field expectation values. For this, let us rewrite 
\begin{equation}
	T_\vk = \left(\begin{array}{cc}
	U_\vk & V_\vk\\ V_\vk & U_\vk	
\end{array}
\right),
\end{equation}
which can always be done as long as the assumptions in \eqnref{eq:Ham_M_3} are true. Since $T_{\vk}^{-1}a_\vk = b_\vk$, we have
\begin{equation}
	a_{i,\vk}=e_i^T T_\vk b_\vk = (U_\vk)_{ij}b_{i,\vk} +(V_\vk)_{ij}b_{j,-\vk}^\dagger,
\end{equation}
such that
\begin{align}
\label{eq:exp_val_1}
	n_i =& \expval*{a_i^\dagger a_i}=\frac{\Nuc}{N}\sum_\vk (V_\vk V_\vk^\dagger)_{ii}\\
\label{eq:exp_val_2}
	m_{ij}=&\expval{a_i^\dagger a_j}=\frac{1}{N}\sum_\vk \gamma_\vk^{(ij)} (V_\vk V_\vk^\dagger)_{ji}\\
\label{eq:exp_val_3}
	\delta_{i}=&\expval{a_i^2}=\frac{\Nuc}{N}\sum_\vk (V_\vk U_\vk^\dagger)_{ii}\\
\label{eq:exp_val_4}
	\Delta_{ij}=&\expval{a_i a_j}=\frac{1}{N}\sum_\vk \gamma_\vk^{(ij)} (V_\vk U_\vk^\dagger)_{ji},
\end{align}
where
\begin{equation}
	\gamma_\vk^{(ij)} = \sum_{\mathbf{d}}e^{i\vk \cdot \mathbf{d}},
\end{equation}
with $\mathbf{d}$ being the set of lattice vectors between sites $ij$ at given distance (NN or NNN).

\subsubsection{Reality of Mean-Field Expectation Values}

Let us quickly prove that the expectation values in Eqs.~(\ref{eq:exp_val_1}-\ref{eq:exp_val_4}) are real. To do so, we assume that \eqnref{eq:Ham_M_2} holds for the matrix that we diagonalize. Then by reality of $\omega_\vk$
\begin{equation}
	T_\vk^\dagger M_\vk T_\vk =  T_\vk^T M_\vk^\ast T_\vk^\ast = T_\vk^T M_{-\vk} T_\vk^\ast,
\end{equation}
we know that $T_{-\vk}=T_\vk^\ast$. As clearly $(\gamma_\vk^{(ij)})^\ast=\gamma_{-\vk}^{(ij)}$, for the MF-expectation values
\begin{equation}
	m_{ij}^\ast =\frac{1}{N}\sum_\vk \gamma_\vk^{(ij)\ast} (V_\vk V_\vk^\dagger)_{ji}^\ast = \frac{1}{N}\sum_\vk \gamma_{-\vk}^{(ij)} (V_{-\vk} V_{-\vk}^\dagger)_{ji}.
\end{equation}
Using the fact that the sum is symmetric around $\vk = 0$, we thus know that $m_{ij}^\ast = m_{ij}$ and similarly $n_i, \delta_i, \Delta_{ij}\in \mathbb{R}$, hence the statement of \eqnref{eq:Ham_M_3} remains true at order $S^{-2}$.

\section{Ground State Algorithm}

Lastly, let us describe the algorithm used to obtaining the proper, self-consistent ground state. That is, we want to determine the state at wich the mean-field expectation values computed from the diagonalization of the matrix $M_\vk$ are identical to the values plugged in.\par
Formally, this is possible analytically though not feasible. We will thus restrict ourselves to a finite lattice with PBC and approximate the self-consistent solution. Since some regions are classically unstable, we have two choices left, which will be discussed in the following subsections. For both, we normalize $J_1=1$.

\subsection{Local Algorithm}

The local algorithm works at a single point in parameter space $J_2, H$ at a given finite, periodic lattice.
\begin{enumerate}
	\item Diagonalize $M_\vk$ at order $S^{-1}$ at all allowed $\vk$.
	\item Compute ground state energy corrections and mean-field expectatino values $F$.
	\item Define new mean-field expectation values $F_\text{new}=\sin^2(\theta)F$.
	\item Diagonalize $M_\vk$ for all $\vk$ and compute energy and mean-field expectation values $F$.
	\item Define new mean-field expectation values by a mixing angle $F_\text{new}=\sin^2(\theta)F_\text{old}+\cos^2(\theta)F$.
	\item Repeat from step 4 until energy has converged or maximum amount of recursions is reached.
\end{enumerate}
As a convergence condition, we assume a Cauchy-sequence like criterion, i.e., if for the last ten recursions, $\abs{\delta E_i-\delta E_j}\leq 10^{-9} \forall i,j$.

\subsection{Stable Algorithm}

Since some configurations are classically unstable but may be stabilized by quantum corrections, we may also use a different algorithm.\par
First, the initial mean-field expectation values and energy is evaluated at a known stable point, $J_{2,\text{i}}, H_{\text{i}}$ using the local algorithm. Then, along a chosen path in parameter space $J_2(t), H(t)$ to the final parameters $J_{2,\text{f}},H_{\text{f}}$ at a given step size $\Delta J_2, \Delta_H$. At each step, the previous set of mean-field parameters is used as input, and the new set of parameters is evaluated until the final point is reached. At this point, step 4 to 6 of the local algorithm are performed.


Note that there is one caveat to each of the algorithms, resolved by the respective other. The local algorithm may become unstable due to a classical instability despite there being a self-consistent solution, while the stable algorithm may reside in a state that is not the ground state.

\onecolumngrid
\section{States and Angle Renormalization}

Angle renormalization follows the scheme that
\begin{equation}
	\Ham^{(1)}+\sum_{n=1}^\infty \overline{\Ham}^{(2n+1)}(\Theta)=0.
\end{equation}
It is sufficient to consider this equation only for the annihilation operators as $\Ham$ is Hermitian. 

\subsection{Three-Site Unit Cell}

In the three-sublattice phase, we need the mean-field parameters
\begin{equation}
	n_i = \expval*{a_i^\dagger a_i},\quad \delta_i = \expval{a_ia_i},\quad m_{ij}=\expval{a_i^\dagger a_j}, \quad \Delta_{ij}\expval{a_i a_j}, \quad \eta_i = \expval{a_i^\dagger a_{i+\mathbf{d}_{\text{NNN}}}},\quad \lambda_i = \expval{a_i a_{i+\mathbf{d}_{\text{NNN}}}},
\end{equation}
where $i,j$ are nearest neighbors. Furthermore, the NNN interaction does not contribute to renormalizing the angles. The general expressions read
\begin{equation}
\begin{split}
	0=&H\sin(\theta_i)\left[1-\frac{1}{4S}(2n_i+\delta_i^\ast)\right]+3J_1\sum_{j\neq i}\sin(\theta_i-\theta_j)\left[-1+\frac{1}{4S}\left(2n_i+\delta_i^\ast-4m_{ij}-4\Delta_{ij}+4n_j\right)\right].
\end{split}
\end{equation}
When constraining to the states defined \cref{fig:three_sublat_states}. Note that this also influences the mean-field expectation values. For example, for the Y-state, we may replace $m_{13}\rightarrow m_{12}$ since these are average quantities and the angle between $1$ and $3$ is the same as between $1$ and $2$. \par
In this manner, we can simplify the different expressions. For all three-sublattice phases in this work, the mean-field parameter relationships are
\begin{equation}
	n_3 = n_2,\quad \eta_3 = \eta_2, \quad \lambda_3 = \lambda_2, \quad m_{13} = m_{12}, \quad \Delta_{13} = \Delta_{12}.
\end{equation}
\subsubsection{Y-State}
Renormalized Angles:
\begin{align}
	\cos(\overline{\theta}^{\text{Y}}) =& \cos(\theta^{\text{Y}})+\frac{1}{S}\left[\cos(\theta^{\text{Y}})(m_{23}-n_2+\Delta_{23})+\frac{1}{2}(m_{21}-n_1+\Delta_{21})\right].
\end{align}
with $\cos(\theta)=\frac{B+3J_1}{6J_1}$ is the classical angle.
\subsubsection{V-State}
Renormalized Angles
\begin{align}
	\cos(\overline{\theta}_1^\text{V}) =&\cos(\theta_1^\text{V})-\frac{1}{S}\left[2\cos(\theta_2^\text{V})(m_{12}+\Delta_{12})+\cos(\theta_1^\text{V})n_1-\frac{3J_1}{H}(n_1-4n_2)\right]\\
	\cos(\overline{\theta}_2^\text{V}) =&\cos(\theta_2^\text{V})-\frac{1}{S}\left[\frac{1}{2}\cos(\theta_1^\text{V})(m_{12}+\Delta_{12})+\cos(\theta_2^\text{V})n_2+\frac{3J_1}{2H}(n_1-4n_2)\right]
\end{align}
with 
\begin{equation}
	 \cos(\theta_1^{\text{V}})=\frac{H^2-27J_1^2}{6J_1H}\quad \cos(\theta_2^{\text{V}}) = \frac{H^2+27J_1^2}{12J_1H}.
\end{equation}
 
\subsection{Four-Sublattice Phase}

In the four-sublattice phase, we need the mean-field parameters
\begin{equation}
	n_i = \expval*{a_i^\dagger a_i},\quad \delta_i = \expval{a_ia_i},\quad m_{ij}^{(l)}=\expval{a_i^\dagger a_j}, \quad \Delta_{ij}^{(l)}\expval{a_i a_j},\quad \overline{m}_{i,j}=m_{i,j}^{(1)}+\frac{J_2}{J_1}m_{i,j}^{(2)},\quad \overline{\Delta}_{i,j}=\Delta_{i,j}^{(1)}+\frac{J_2}{J_1}\Delta_{i,j}^{(2)},
\end{equation}
where $i,j$ are nearest ($l=1$) or next-nearest ($l=2$) neighbors. Furthermore, we shall use $\overline{J}\equiv J_1+J_2$.

\subsubsection{Canted Stripe State}

For the canted stripe state, we have
\begin{equation}
	\cos(\overline{\theta})=\cos(\theta)+\frac{\cos(\theta)}{S}\left[n_1-\frac{J_1}{\overline{J}}\left(\overline{m}_{13}+\overline{\Delta}_{13}\right)\right],\quad \text{for}\quad \cos(\theta)=\frac{H}{8\overline{J}}.
\end{equation}

\subsubsection{$\downarrow \times$V State}

We use the result from the V state renormalization to find
\begin{align}
	\cos(\overline{\theta}_2)=&\cos(\theta_2)+\frac{1}{S}\left[\frac{4\overline{J}\cos(\theta_3)}{H+2\overline{J}}\left(\frac{J_1}{\overline{J}}\left(\overline{m}_{13}+\overline{\Delta}_{13}\right)-n_1\right)-\frac{16\overline{J}J_1}{(H+2\overline{J})^2}\left(\overline{\Delta}_{13}+\overline{m}_{13}-\overline{\Delta}_{12}-\overline{m}_{12}\right)\right.\\&\left.-\frac{2J_1\cos(\theta_3)}{\overline{J}}\left(\overline{m}_{23}+\overline{\Delta}_{23}\right)+\cos(\theta_2)n_2+\frac{2\overline{J}}{H+2\overline{J}}(4n_3-n_2)\right]\\
		\cos(\overline{\theta}_3)=&\cos(\theta_3)+\frac{1}{S}\left[\frac{\overline{J}\cos(\theta_2)}{H+2J}\left(\frac{J_1}{\overline{J}}\left(\overline{m}_{12}+\overline{\Delta}_{12}\right)-n_1\right)+\frac{2\overline{J}J_1}{(H+2\overline{J})^2}\left(\overline{\Delta}_{13}+\overline{m}_{13}-\overline{\Delta}_{12}-\overline{m}_{12}\right)\right.\\&\left.-\frac{J_1\cos(\theta_2)}{2\overline{J}}\left(\overline{m}_{23}+\overline{\Delta}_{23}\right)+\cos(\theta_3)n_3-\frac{\overline{J}}{H+2\overline{J}}(4n_3-n_2)\right]
\end{align}
where 
\begin{equation}
	\cos(\theta_2)=\frac{H^2+4H\overline{J}-8\overline{J}^2}{4\overline{J}(H+2\overline{J})},\quad \cos(\theta_3)=\frac{H^2+4H\overline{J}+16\overline{J}^2}{8\overline{J}(H+2\overline{J})}.
\end{equation}

\subsubsection{$\overline{\text{V}}$ State}

We again use the results from the V state computation to obtain
\begin{align}
	\cos(\overline{\theta}_1)=&\cos(\theta_1)+\frac{1}{S}\left[-\frac{3J_1}{\overline{J}}\cos(\theta_2)\left(\overline{m}_{12}+\overline{\Delta}_{12}\right)+\cos(\theta_1)n_1+\frac{2\overline{J}}{H}\left(9n_2-n_1\right)\right]\\
	\cos(\overline{\theta}_2)=&\cos(\theta_2)+\frac{1}{S}\left[-\frac{J_1\cos(\theta_1)}{3\overline{J}}\left(\overline{m}_{12}+\overline{\Delta}_{12}\right)+\cos(\theta_2)n_2-\frac{2\overline{J}}{3H}\left(9n_2-n_1\right)\right]\\
\end{align}
where 
\begin{equation}
	\cos(\theta_1)=\frac{H^2-32\overline{J}^2}{4HJ},\quad \cos(\theta_2)=\frac{H^2+32\overline{J}^2}{12HJ}.
\end{equation}

\section{Expansion in $S^{-1}$}

For the self-consistent mean-field treatment, we will need the expansion of the terms $\frac{\vec{S}_i \cdot \vec{S}_j}{S^2}$ and $\frac{\vec{B}\cdot \vec{S}_i}{S}$ to order $S^{-5/2}$. We shall use the expansion
\begin{equation}
	S_i^+ = \sqrt{2S-n_i}a_i = \sqrt{2S}a_i-\frac{1}{2\sqrt{2S}}n_ia_i-\frac{1}{16\sqrt{2}S^{3/2}}n_i^2a_i+\mathcal{O}\left(S^{-5/2}\right).
\end{equation}
Using this, we find the following contributions to order $S^{-5/2}$:
\begin{align}
	\frac{S_i^+S_j^+}{S^2}=&\frac{1}{S}a_ia_j-\frac{1}{2S^2}\left(n_ia_ia_j+a_in_ja_j\right)+\mathcal{O}(S^{-3})\\
	\frac{S_i^+S_j^-}{S^2}=&\frac{1}{S}a_ia_j^\dagger-\frac{1}{2S^2}\left(n_ia_ia_j^\dagger+a_ia_j^\dagger n_j\right)+\mathcal{O}(S^{-3})\\
	\frac{S_i^+S_j^z}{S^2}=&\sqrt{\frac{2}{S}}a_i-\frac{1}{2\sqrt{2}S^{3/2}}\left(n_ia_i+4a_in_j\right)-\frac{1}{16\sqrt{2}S^{5/2}}\left(n_i^2a_i-8n_ia_in_j\right)+\mathcal{O}(S^{-7/2}),\\
	\frac{S_i^zS_j^z}{S^2}=&1-\frac{1}{S}(n_i+n_j)+\frac{1}{S^2}n_in_j.
\end{align}
Defining the mean-field expectation values
\begin{equation}
	F_{ij}=\expval{a_i^\dagger a_j}, \quad \Delta_{ij}=\expval{a_i a_j},\quad n_i = F_{ii}, \quad \delta_i = \Delta_{ii},
\end{equation}
we may reduce the above expressions to the point they only contain constant terms as well as terms linear and quadratic in the bosonic operators. When considering the spins in a coplanar configuration, we have the expression in the lab frame given as
\begin{equation}
	\left(\begin{array}{c}
		S_l^x \\ S_l^y \\ S_l^z 
	\end{array}\right)= \left(\begin{array}{c}
	\cos(\theta) {S}^x + \sin(\theta){S'}^z\\ {S}^y \\\cos(\theta){S}^z-\sin(\theta){S}^x 
	\end{array}\right),
\end{equation}
we find
\begin{align}
	\frac{\vec{S}_i \cdot \vec{S}_j}{S^2}=&\cos(\theta_i-\theta_j)+\frac{\sin(\theta_i-\theta_j)}{\sqrt{2S}}\left[a_j + a_j^\dagger -a_i-a_i^\dagger\right]\\
	&+\frac{1}{S}\left[\frac{\cos(\theta_i-\theta_j)-1}{2}(a_i^\dagger a_j^\dagger+a_ia_j) -\cos(\theta_i-\theta_j)(a_i^\dagger a_i+a_j^\dagger a_j)+\frac{\cos(\theta_i-\theta_j)+1}{2}(a_i^\dagger a_j+a_j^\dagger a_i)\right]\\
	&+\frac{\sin(\theta_i-\theta_j)}{4\sqrt{2}S^{3/2}}\left[a_i \left(2 n_i+4 n_j-4\Delta _{ij}^*-4 m_{ij}+\delta _i^*\right)+a_i^{\dagger } \left(2 n_i+4 n_j-4 \Delta _{ij}-4 m_{ij}^\ast+\delta _i\right)\right.\\&\left.-a_j \left(2n_j+4n_i- 4\Delta _{ij}^*-4m_{ij}^\ast+\delta _j^*\right)-a_j^{\dagger } \left(2n_j+4n_i-4 \Delta _{ij}-4 m_{i,j}+\delta _j\right)\right]\\
	&+\frac{1}{S^2}\left[-\left(\frac{\cos(\theta_i-\theta_j)+1}{8}\Delta_{ij}+\frac{\cos(\theta_i-\theta_j)-1}{8}F_{ij}^\ast\right)(a_i^\dagger)^2\right.\\
	&-\left.\left(\frac{\cos(\theta_i-\theta_j)+1}{8}\Delta_{ij}+\frac{\cos(\theta_i-\theta_j)-1}{8}F_{ij}\right)(a_j^\dagger)^2\right.\\
	&+\left.\left(\cos(\theta_i-\theta_j)\Delta_{ij}-\frac{\cos(\theta_i-\theta_j)-1}{4}(n_i+n_j)-\frac{\cos(\theta_i-\theta_j)+1}{8}(\delta_i+\delta_j)\right)a_i^\dagger a_j^\dagger\right.\\
	&+\left.\left(\cos(\theta_i-\theta_j)n_j-\frac{\cos(\theta_i-\theta_j)+1}{2}\Re{F_{ij}}-\frac{\cos(\theta_i-\theta_j)-1}{2}\Re{\Delta_{ij}}\right)a_i^\dagger a_i\right.\\
	&+\left.\left(\cos(\theta_i-\theta_j)F_{ij}^\ast-\frac{\cos(\theta_i-\theta_j)+1}{4}(n_i+n_j)-\frac{\cos(\theta_i-\theta_j)-1}{8}(\delta_i+\delta_j^\ast)\right)a_i^\dagger a_j\right.\\
	&+\left.\left(\cos(\theta_i-\theta_j)F_{ij}-\frac{\cos(\theta_i-\theta_j)+1}{4}(n_i+n_j)-\frac{\cos(\theta_i-\theta_j)-1}{8}(\delta_i^\ast+\delta_j)\right)a_j^\dagger a_i\right.\\
	&+\left.\left(\cos(\theta_i-\theta_j)n_i-\frac{\cos(\theta_i-\theta_j)+1}{2}\Re{F_{ij}}-\frac{\cos(\theta_i-\theta_j)-1}{2}\Re{\Delta_{ij}}\right)a_j^\dagger a_j\right.\\
	&-\left.\left(\frac{\cos(\theta_i-\theta_j)+1}{8}\Delta_{ij}^\ast+\frac{\cos(\theta_i-\theta_j)-1}{8}F_{ij}\right)(a_i)^2\right.\\
	&-\left.\left(\frac{\cos(\theta_i-\theta_j)+1}{8}\Delta_{ij}^\ast+\frac{\cos(\theta_i-\theta_j)-1}{8}F_{ij}^\ast\right)(a_j)^2\right.\\
	&+\left.\left(\cos(\theta_i-\theta_j)\Delta_{ij}^\ast-\frac{\cos(\theta_i-\theta_j)-1}{4}(n_i+n_j)-\frac{\cos(\theta_i-\theta_j)+1}{8}(\delta_i^\ast+\delta_j^\ast)\right)a_i^\dagger a_j^\dagger\right.\\
	&-\left.\cos(\theta_i-\theta_j)\left(\abs{F_{ij}}^2+\abs{\Delta_{ij}}^2+n_in_j\right)+\frac{\cos(\theta_i-\theta_j)+1}{4}\Re\left\{(\delta_i+\delta_j)\Delta_{ij}^\ast+2(n_i+n_j)F_{ij}\right\}\right.\\
	&+\left.\frac{\cos(\theta_i-\theta_j)-1}{4}\Re\left\{\delta_i F_{ij}+\delta_j F_{ij}^\ast+2(n_i+n_j)\Delta_{ij}\right\}\right]+\mathcal{O}(S^{-5/2}).
\end{align}
and
\begin{align}
	\frac{\vec{B} \cdot \vec{S}_i}{S}=&-B\cos(\theta_i)+\frac{B\sin(\theta_i)}{\sqrt{2S}}(a_i+a_i^\dagger)+\frac{B\cos(\theta_i)}{S}a_i^\dagger a_i+\frac{B\sin(\theta_i)}{4\sqrt{2}S^{3/2}}\left[(\delta_i^\ast +2n_i)a_i+(\delta_i +2n_i)a_i^\dagger\right]+\mathcal{O}(S^{-5/2})
\end{align}
for $B=(0,0,-B)^T$.\par
From the above expressions, we may read off the corrections to the matrices.

\subsection{Three-Sublattice Phase}

The matrix for the bosonic operators $a_\vk = (a_{1,\vk},a_{2,\vk},a_{3,\vk},a_{1,-\vk}^\dagger,a_{2,-\vk}^\dagger,a_{3,-\vk}^\dagger)^T$, following the notation in 
\eqnref{eq:Ham_M_2} reads
\begin{align}
	A_{ii}=&-3J_1\sum_{j\neq i}\cos(\theta_i-\theta_j)-6J_2(1-\Gamma_\vk)+H\cos(\theta_i)+\frac{1}{S}\left[6J_2(n_i-\eta_i)(1-\Gamma_\vk)+3J_1\sum_{j\neq i}(\delta \theta_i-\delta \theta_j)\sin(\theta_i-\theta_j)\right.\\
	&-\left.H\delta\theta_i\sin(\theta_i)+3J_1 \sum_{j\neq i}\left(\cos(\theta_i-\theta_j)n_j-\frac{\cos(\theta_i-\theta_j)+1}{2}\Re{m_{ij}}-\frac{\cos(\theta_i-\theta_j)-1}{2}\Re{\Delta_{ij}}\right)\right]\\
	A_{ij}=&J_1\gamma_\vk^{(\ast)}\left(\frac{\cos(\theta_i-\theta_j)+1}{2}+\frac{1}{S}\left[\cos(\theta_i-\theta_j)F_{ij}^\ast-\frac{\cos(\theta_i-\theta_j)+1}{4}(n_i+n_j)-\frac{\cos(\theta_i-\theta_j)-1}{8}(\delta_i+\delta_j^\ast)\right.\right.\\&\left.\left.-\frac{\delta \theta_i-\delta \theta_j}{2}\sin(\theta_i-\theta_j)\right]\right)\\
	B_{ii}=&\frac{1}{S}\left[-3J_1\sum_{j\neq i}\left(\frac{\cos(\theta_i-\theta_j)+1}{8}\Delta_{ij}+\frac{\cos(\theta_i-\theta_j)-1}{8}F_{ij}^\ast\right)+ 6J_2 \left(-\frac{\lambda_i}{4}+\frac{\Gamma_\vk}{2}(2\lambda_i-\delta_i)\right)\right]\\
	B_{ij}=&J_1\gamma_\vk^{(\ast)}\left(\frac{\cos(\theta_i-\theta_j)-1}{2}+\frac{1}{S}\left[\cos(\theta_i-\theta_j)\Delta_{ij}-\frac{\cos(\theta_i-\theta_j)-1}{4}(n_i+n_j)\right.\right.\\&\left.\left.-\frac{\delta \theta_i-\delta \theta_j}{2}\sin(\theta_i-\theta_j)-\frac{\cos(\theta_i-\theta_j)+1}{8}(\delta_i+\delta_j)\right]\right)
\end{align}
for
\begin{equation}
	\gamma_\vk = e^{i\vk \cdot \mathbf{a}_1}+e^{-i\vk \cdot \mathbf{a}_2}+e^{i\vk \cdot (\mathbf{a}_2-\mathbf{a}_1)}, \quad 3\Gamma_\vk = \cos(\vk \cdot (\mathbf{a}_1+\mathbf{a}_2))+\cos(\vk \cdot (2\mathbf{a}_1-\mathbf{a}_2))+\cos(\vk \cdot (-\mathbf{a}_1+2\mathbf{a}_2))
\end{equation}
and
\begin{equation}
	\overline{\theta_i}=\theta_i+\frac{1}{S}\delta \theta_i+\mathcal{O}(S^{-2}).
\end{equation}
There direct energy correction from the mean-field approximation reads
\begin{align}
	\delta E_{\text{MF}}^{(4)}=&3J_1 \sum_{i,j\neq i}\left[-\cos(\theta_i-\theta_j)\left(\abs{F_{ij}}^2+\abs{\Delta_{ij}}^2+n_in_j\right)+\frac{\cos(\theta_i-\theta_j)+1}{4}\Re\left\{(\delta_i+\delta_j)\Delta_{ij}^\ast+2(n_i+n_j)F_{ij}\right\}\right.\\
	&+\left.\frac{\cos(\theta_i-\theta_j)-1}{4}\Re\left\{\delta_i F_{ij}+\delta_j F_{ij}^\ast+2(n_i+n_j)\Delta_{ij}\right\}\right]+6J_2\sum_i \left[-(\eta_i^2+\lambda_i^2+n_i^2)+\Re{\delta_i \lambda_i^\ast +2n_i \eta_i}\right].
\end{align}

\subsection{Four-Sublattice Phase}

Similarly, the matrix elements in the four-sublattice phase are given by

\begin{align}
	A_{ii}=&-2\overline{J}\sum_{j\neq i}\cos(\theta_i-\theta_j)+H\cos(\theta_i)+\frac{1}{S}\left[2\overline{J}\sum_{j\neq i}(\delta \theta_i-\delta \theta_j)\sin(\theta_i-\theta_j)-H\delta\theta_i\sin(\theta_i)\right.\\
	&\left.+ \sum_{l,j\neq i}2J_l\left(\cos(\theta_i-\theta_j)n_j-\frac{\cos(\theta_i-\theta_j)+1}{2}\Re{m_{ij}^{(l)}}-\frac{\cos(\theta_i-\theta_j)-1}{2}\Re{\Delta_{ij}^{(l)}}\right)\right]\\
	A_{ij}=&\sum_l J_l\gamma_\vk^{(ij, l)}\left(\frac{\cos(\theta_i-\theta_j)+1}{2}+\frac{1}{S}\left[\cos(\theta_i-\theta_j)(m_{ij}^{(l)})^\ast-\frac{\cos(\theta_i-\theta_j)+1}{4}(n_i+n_j)\right.\right.\\&\left.\left.-\frac{\cos(\theta_i-\theta_j)-1}{8}(\delta_i+\delta_j^\ast)-\frac{\delta \theta_i-\delta \theta_j}{2}\sin(\theta_i-\theta_j)\right]\right)\\
	B_{ii}=&\frac{1}{S}\left[\sum_{j\neq i}-2J_l\left(\frac{\cos(\theta_i-\theta_j)-1}{8}m_{ij}^{(l)}+\frac{\cos(\theta_i-\theta_j)+1}{8}\Delta_{ij}^{(l)}\right)\right]\\
	B_{ij}=&\sum_l J_l\gamma_\vk^{(ij,l)}\left(\frac{\cos(\theta_i-\theta_j)-1}{2}+\frac{1}{S}\left[\cos(\theta_i-\theta_j)\Delta_{ij}^{(l)}-\frac{\cos(\theta_i-\theta_j)-1}{4}(n_i+n_j)\right.\right.\\&\left.\left.-\frac{\delta \theta_i-\delta \theta_j}{2}\sin(\theta_i-\theta_j)-\frac{\cos(\theta_i-\theta_j)+1}{8}(\delta_i+\delta_j)\right]\right)
\end{align}
for
\begin{equation}
	\overline{J}=J_1+J_2, \quad \gamma_\vk^{(ij,l)}=\sum_{\mathbf{d}}2\cos(\mathbf{d}_{ij}\cdot \vk), \quad \overline{\theta_i}=\theta_i+\frac{1}{S}\delta \theta_i+\mathcal{O}(S^{-2}).
\end{equation}
The direct mean-field correction is given by
\begin{align}
	\delta E_{\text{MF}}^{(4)}=&\sum_{i\neq j}J_l\left[-\cos(\theta_i-\theta_j)\left(\abs{m_{ij}^{(l)}}^2+\abs{\Delta_{ij}^{(l)}}^2+n_in_j\right)+\frac{\cos(\theta_i-\theta_j)+1}{4}\Re\left\{(\delta_i+\delta_j)(\Delta_{ij}^{(l)})^\ast+2(n_i+n_j)m_{ij}^{(l)}\right\}\right.\\
	&+\left.\frac{\cos(\theta_i-\theta_j)-1}{4}\Re\left\{\delta_i m_{ij}^{(l)}+\delta_j (m_{ij}^{(l)})^\ast+2(n_i+n_j)\Delta_{ij}^{(l)}\right\}\right].
\end{align}

\section{Transition to the fully polarized phase}
\maketitle
Here we derive the transition magnetic field between the fully polarized phase and the V phase for the $J_1 - J_2 -h$ model. This transition can be computed exactly using the global spin conservation. 
The spin-$\frac{1}{2}$ Hamiltonian is defined as 
\begin{equation}
	H = H_0 - h \sum_i S_i^z \equiv J_1 \sum_{\expval{ij}} \mathbf{S}_i \cdot \mathbf{S}_j + J_2\sum_{\expval{\expval{ik}}} \mathbf{S}_i \cdot \mathbf{S}_k - h \sum_i S_i^z
\end{equation}
where the double braket denotes next-nearest neighbor bonds. Again the Hamiltonian conserves $S_{\rm tot}^z$ thus we can apply the trick in the previous section. 
The critical field for polarization is the lowest magnetic field at which the system becomes fully saturated in magnetization, i.e. $S^z_{\rm tot} = N/2 = L^2/2$. For this to happen, the $S^z_{\rm tot} = N/2-1$ sector must have lower  energy than the sector with $S_{\rm tot}^z = N/2$, thus the critical field $h_c$ satisfies
\begin{equation} \label{eq:eheh}
	E_0(\frac{N}{2}) - h_c \times \frac{N}{2} = E_0(\frac{N}{2}-1) - h_c\times(\frac{N}{2} - 1)
\end{equation}
where $E_0(N)$ is a short notation for the energy in the $S^z_{\rm tot} = N$ sector without magnetic field (of $H_0$). Note that the $S_{\rm tot}^z = N/2$ sector has only one state, with all spins pointing along $S^z = \frac{1}{2}$, and the total energy is given by 
\begin{equation}
	E_0(N/2) = H_0 \ket*{S_{\rm tot}^z = \frac{N}{2}} = 3N J_1 \frac{1}{4} + 3N J_2 \frac{1}{4} = \frac{3}{4}N (J_1 + J_2)
\end{equation}
since there are three bonds per site. 
While the sector with $S_{\rm tot}^z = N/2-1$ has exactly $N$ states, each of which has $N/2-1$ spins pointing in $S^z = \frac{1}{2}$ and one spin pointing in $S^z = -\frac{1}{2}$. Next we need to find the lowest energy in this sector, let us denote the state with $S_n^z = -\frac{1}{2}$ by $\ket{f_n}$, and consider the energy function of such a state. 
\begin{equation}
	\begin{split}
		J_1&\sum_{i,j\in\{a_1, a_2, a_3\}}\left[S_i^z S_{j}^z + \frac{1}{2} \left( S_i^+ S_{j}^- + S_i^- S_{j}^+ \right)\right] \ket{f_n} = \frac{J_1}{2}\sum_\delta \ket{f_{n\pm \delta}} + (\frac{3N}{4}- 3)J_1\ket{f_n}
	\end{split}
\end{equation}
with $\delta \in \{a_1, a_2, a_3\}$ and $a_3 = a_2 - a_1$.  Now consider the polarizing transition at $J_2 = 0$ only, thus we consider
 \begin{equation}
	 H_{J_1} \ket{f_n} = J_1\left(\frac{3N}{4} - 3\right)\ket{f_n} + \frac{J_1}{2} \sum_\delta \ket{f_{n \pm \delta}}
\end{equation}
Fourier Transform gives
\begin{equation}
	E_0(\frac{N}{2} - 1) \ket{f_k} = J_1 \left[ \left( \frac{3N}{4} - 3 \right) + \sum_{\delta}\cos(k.\delta)\right]
\end{equation}
hence, using $a_1 = (\frac{1}{2}, \sqrt{3}/2)$ and $a_2 = (1,0)$, the lowest energy is
\begin{equation}
	{\rm min}[E_0(N/2-1)] = J_1 \left(\frac{3N}{4} - 4.5\right)
\end{equation}
hence
\begin{equation}
	h_c(J_2 = 0) = 4.5
\end{equation}
This number is veried by DMRG with an error $< 10^{-2}$. 

Next let us consider the generic case for $J_2 \neq 0$. The ground state energy in the two sectors is the same as $J_2 = 0$ case except for a change in the momentum space distribution of the energies. The total energy in the $N/2-1$ sector is thus
\begin{equation}
	E_0(\frac{N}{2} - 1) = (1+ J_2/J_1) \left(\frac{3N}{4} - 3\right) + \sum_\delta \cos(k.\delta) + (J_2/J_1) \sum_{\eta} \cos(k.\eta)
\end{equation}
where $\eta$ are the three next-nearest-neighbor vectors, and we set $J_2$ in units of  $J_1$.  Hence 
\begin{equation}
	{\rm min}[E_0(\frac{N}{2} -1 )] = (1+J_2/J_1) \left( \frac{3N}{4} -3\right) + {\rm min}\left[\sum_\delta \cos(k.\delta) + \frac{J_2}{J_1} \sum_{\eta} \cos(k.\eta) \right]
\end{equation}
This minimum can be readily found for different $J_2/J_1$. The resultant $h_c$ is a piecewise function:
\begin{equation}
	h_c (J_2/J_1) = 
	\begin{cases}
		4.5 & 0< J_2/J_1 \le \frac{1}{8}\\
		  4(1 + J_2/J_1) & J_2/J_1 > \frac{1}{8}
\end{cases}
\end{equation}
where the piece-wise function has a kink exactly at $J_2/J_1 = \frac{1}{8}$ where the Dirac QSL is present at zero field.

\end{document}